\documentclass[manuscript]{acmart}

\AtBeginDocument{%
  }

\usepackage{graphicx}

\usepackage{multirow}
\usepackage{appendix}
\usepackage{pifont}

\usepackage{todonotes}

\usepackage{hyperref}
\usepackage{cleveref}
\hypersetup{breaklinks=true}

\setcopyright{acmlicensed}

\copyrightyear{2025}
\acmYear{2025}
\setcopyright{rightsretained}
\acmConference[MuC '25]{Mensch und Computer 2025}{August 31-September 03, 2025}{Chemnitz, Germany}
\acmBooktitle{Mensch und Computer 2025 (MuC '25), August 31-September 03, 2025, Chemnitz, Germany}
\acmPrice{}
\acmDOI{10.1145/3743049.3748566}
\acmISBN{979-8-4007-1582-2/25/08}
\begin{document}

\title[Evaluation of a Provenance Management Tool]{Evaluation of a Provenance Management Tool for Immersive Virtual Fieldwork}

\author{Armin Bernstetter}
\orcid{0000-0003-1603-1699}
\affiliation{%
  \institution{GEOMAR Helmholtz Centre for Ocean Research Kiel \& Kiel University}
  \city{Kiel}
  \country{Germany}
}
\email{abernstetter@geomar.de}

\author{Tom Kwasnitschka}
\orcid{0000-0003-1046-1604}
\affiliation{%
  \institution{GEOMAR Helmholtz Centre for Ocean Research Kiel}
  \city{Kiel}
  \country{Germany}
}
\email{tkwasnitschka@geomar.de}

\author{Isabella Peters}
\orcid{0000-0001-5840-0806}
\affiliation{%
  \institution{ZBW Leibniz-Information Center for Economics \& Kiel University}
  \city{Kiel}
  \country{Germany}}
\email{ipe@informatik.uni-kiel.de}

\renewcommand{\shortauthors}{Bernstetter et al.}

\begin{abstract}
Ensuring reproducibility of research is an integral part of good scientific practice.
One way to support this is through provenance: information about research workflows from data gathering to researchers' sensemaking processes leading to published results.
This is highly important in disciplines such as geosciences, where researchers use software for interactive and immersive visualizations of geospatial data, doing virtual measurements in simulated fieldwork on 3D models.
We evaluated a provenance management tool, which allows recording of interactions with a virtual fieldwork tool and annotating different states of the visualization.
The user study investigated how researchers used this Digital Lab Book (DLB) and whether perceived ease of use and perceived usefulness differed between groups in immersive or non-immersive settings.
Participants perceived the DLB as both useful and easy to use. While there were indications of differences in perceived ease of use (higher for immersive setting), usage patterns showed no significant group differences.
\end{abstract}




\maketitle

\section{Introduction}
\label{sec:intro}
The reproducibility of scientific results is of utmost importance in many disciplines \cite{fidler2021reproducibility}, especially when researchers engage in complex sensemaking processes \cite{pirolli2005sensemaking}.
The geosciences are no exception \cite{liu2019improving,steventon2022reproducibility}, but they are an interesting case combining personal fieldwork (e.g., visiting an outcrop and recording observations, measurements, and interpretation in handwritten notebooks) with software-based analyses (e.g. using Geographic Information Systems (GIS)).
Virtual fieldwork on digital models of remote or extreme environments introduces digital tools to conduct otherwise impossible measurements and to document interpretations objectively e.g. in the case of photogrammetric seafloor surveying  \cite{kwasnitschka2013doing,arnaubec2023underwater}.

A wide spectrum of extended (XR), virtual reality (VR) or immersive environments, such as domes \cite{kwasnitschka2023spatially}, can be used to improve spatial understanding through a simulated experience of geological structures in their perceived actual size and state \cite{bonali2024geavr, caravaca2020digital,klippel2019transforming, klippel2020value,metois2021oceanic, schuchardt2007benefits,zhao2019harnessing}.
Strong reliance on visual representations of real world properties and a need to personally analyze and manipulate those to arrive at scientific discoveries challenge such traditional methodologies - especially with regard to their reproducibility.
And although visualizations can document a result as a discrete state in the research process, they are volatile in the sense that one needs to understand how they were generated to provide scientific merit \cite{vanwijk2005value} and to allow for collaborative and reproducible research.
This requires researchers to document their work and decision-making process and to provide provenance information \cite{moreau2011provenance} - or better: the GIS they use should take care of this \cite{ziegler2023needfinding}.

This paper presents the evaluation of a tool developed for this purpose \cite{bernstetter2023practical}. 
This "Digital Lab Book" (DLB) provides provenance management for research in a virtual fieldwork tool (VFT) \cite{bernstetter2025virtualfieldwork}.
Both applications' source code is available (see section \nameref{subsec:code}).
We investigate how researchers make use of the DLB to aid reproducibility of their research process and we study whether the setting (immersive vs. non-immersive) impacts usage behavior, perceived usefulness and perceived ease of use of the DLB. 
For this we provide a simple setting with a desktop PC as well as a spatially immersive projection dome, the ARENA2\footnote{\url{https://www.geomar.de/en/arena}} at GEOMAR Helmholtz Centre for Ocean Research Kiel \cite{kwasnitschka2023spatially}.

\begin{figure*}[!htbp]
    \centering
    \includegraphics[width=\linewidth]{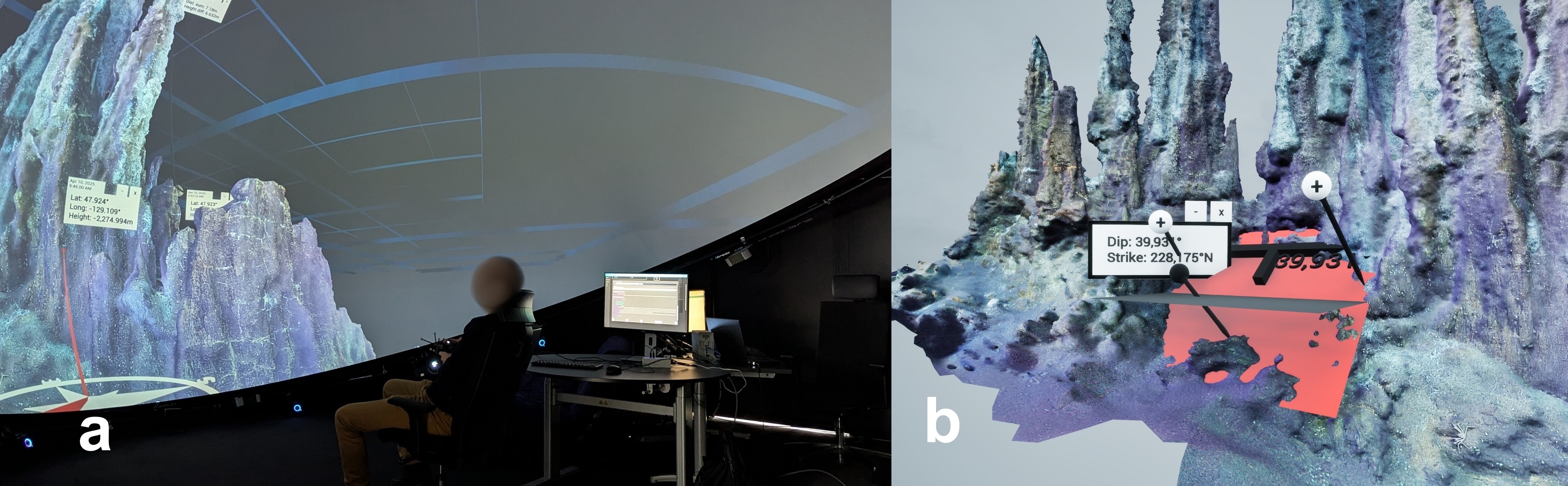}
    \caption{\textbf{a)} The VFT and DLB being used in the "Dome" setting. Participants were sitting in the center of the area below the tilted (21°) dome. On a table right next to the participant, the DLB could be used on a standard PC monitor. \textbf{b)} A closer look at a 3D model showing hydrothermal vents and a Strike \& Dip measurement, a geological convention to measure the angle of a slope.}
    \label{fig:vft_dome_setup}
    \Description{Left: A person sitting in front of and under a large multi-projection dome showing a visualized 3D model of a hydrothermal vent. Right: The same hydrothermal vent 3D model including a virtual measurement of a slope angle}
\end{figure*}
\section{Background \& Related Work}
\label{sec:bg-rel}
"Provenance" can be defined as "information about entities, activities, and people involved in producing a piece of data or thing, which can be used to form assessments about its quality, reliability or trustworthiness" \cite{moreau2011provenance,2013provoverview}.
This includes data provenance (changes to data over time, e.g. in a database) but also interaction history (how users interact with a software) and knowledge provenance (e.g. scientists' thought processes) \cite{fekete2019provenance, ragan2015evaluating,xu2020survey}. 
Provenance provides researchers with information on work done in the past which in turn can provide help for solving current problems \cite{wexelblat1999footprints}.
Provenance management can itself be part of a sensemaking process \cite{shrinivasan2008supporting,wexelblat1999footprints} and help "assess the credibility of hypotheses based on the origins of the supporting information" \cite{toniolo2015supporting}. 

Several approaches have been made to allow for recording of provenance information.
Similar to the DLB \citet{frew2001earth} introduced a provenance system acting as a virtual lab notebook built with a client-server-architecture.
\citet{delrio2007identifying,delrio2007probeit} employ a knowledge provenance system to find imperfections in maps generated in a GIS and \citet{howe2008endtoend} employ VisTrails \cite{callahan2006vistrails} for provenance management at an ocean observatory.
They, as well as \citet{anderson2006visualization}, also use VisTrails for the provenance of interactive 3D data visualization, the latter aiming towards "replacing the laboratory notebook" for the analysis of medical data.

Our study aims to fill two research gaps: 
\begin{enumerate}
    \item It was found that users of geospatial data are missing provenance information in GIS \cite{ziegler2023needfinding} and that systems for immersive interactive geospatial data visualization (\cite{bonali2024geavr, caravaca2020digital,klippel2019transforming, klippel2020value,metois2021oceanic, zhao2019harnessing}) are not yet equipped with provenance management capabilities.
    \item Although \citet{zhang2023defining} have introduced the concept of "embodied provenance" to study the support of provenance in immersive analytics (IA)\cite{marriott2018immersive}, there is a lack of research on provenance management for IA in the geosciences.
\end{enumerate}
While our focus lies on investigating and improving the reproducibility of interactive immersive visualization at our ocean research institute, the DLB is completely agnostic to which research field it might be used in.
\section{Methodology: Software \& User Study}
\label{methodology}

\subsection{Virtual Fieldwork Tool \& Digital Lab Book}\label{software}
The VFT is an application built with the game engine Unreal Engine to explore and measure 3D models of geological structures. 
\Cref{fig:vft_dome_setup} shows the VFT in use in a spatially immersive projection dome for immersive scientific visualization. 
The VFT can run standalone but is designed to work with the DLB to enhance scientific value by recording provenance of user actions.
The DLB is a web-based application with its backend running on the same PC as the VFT and operated in a web browser.
Each placement or removal of a measurement (e.g. location markers, distance measurements, slope angles) in the VFT is relevant for the reproducibility of the interaction workflow. 
The DLB uses a history management concept based on the version control system Git to create "provenance repositories".
With these repositories, interactions with the VFT are made persistent, organizable, and publishable/shareable for collaboration and transparency.

\begin{figure*}[!htbp]
    \centering
    \includegraphics[width=\linewidth]{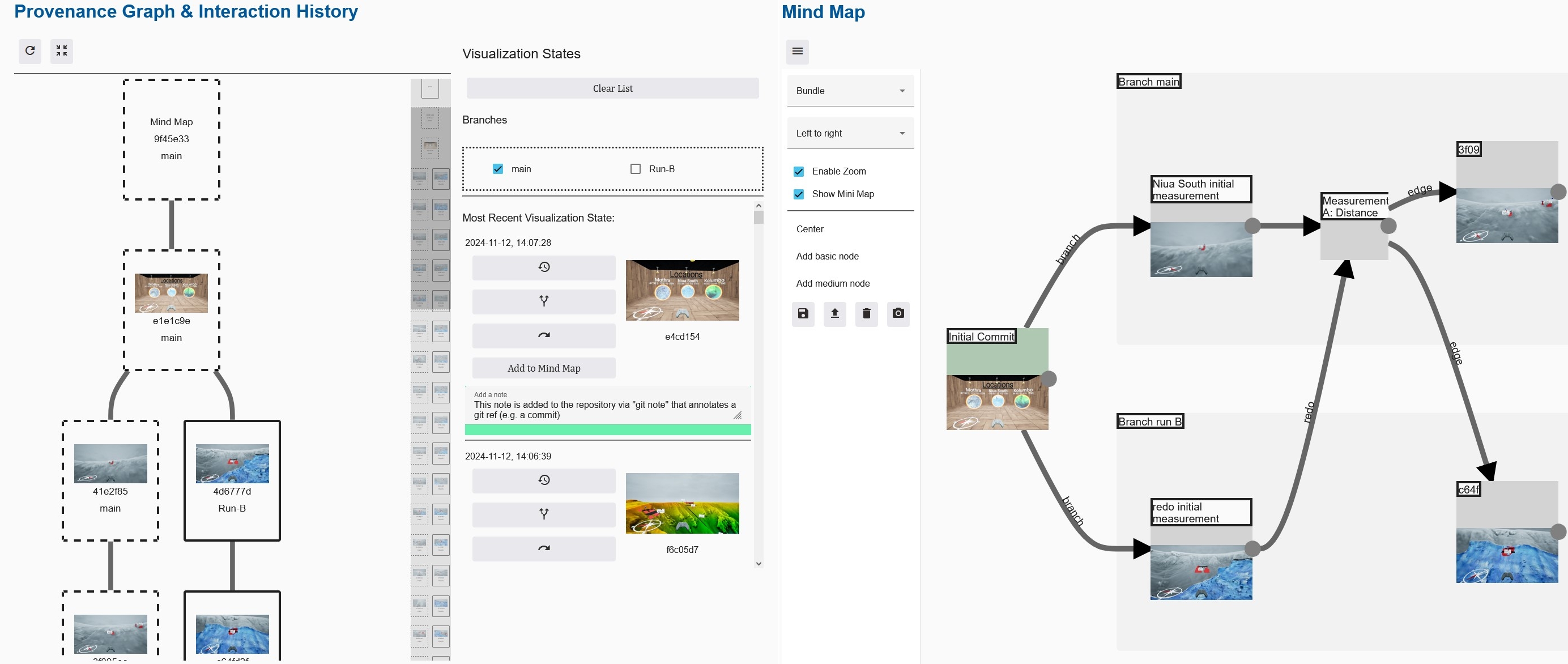}
    \caption{\textbf{Left:} The provenance graph of the DLB showing a visualization of the git branches and elements to annotate visualization states, trigger a return to a state or a branching action. \textbf{Right:} The Mind Map of the DLB showing a selection of commits/visualization states organized and annotated by the user}
    \label{fig:dlb-provgraph-mindmap}
    \Description{Left: The provenance graph of the DLB showing a visualization of the git branches and elements to annotate visualization states, trigger a return to a state or a branching action. Right: The Mind Map of the DLB showing a selection of commits/visualization states organized and annotated by the user}
\end{figure*}

Users can connect the DLB to the VFT and either load an existing repository or create a new one which then saves interaction steps as commits including a screenshot of the VFT in the repository.
Users can annotate or redo actions, restore states, create branches, organize findings in a mind-map, and write additional notes.
Commits from the VFT can also be enriched by speech to text recognition.
In the DLB, a graph visualizes the tree structure of the repository and a mind map allows users to organize, connect, and annotate visualization states (see \cref{fig:dlb-provgraph-mindmap}).
The created and populated repositories can be packaged and exported, and also imported again to the DLB.

A comprehensive introductory video on the usage and interaction patterns between DLB and VFT can be accessed via the DLB's repository\footnote{\url{https://git.geomar.de/digital-lab-book/digital-lab-book\#introduction-video-using-the-digital-lab-book-and-virtual-fieldwork-tool}}.

While the concepts of provenance tracking for supporting reproducibility of scientific results have been investigated e.g. with projects such as VisTrails, the DLB distinguishes itself with its native git integration, allowing easy sharing of provenance repositories as well as external investigation by simply using git on the command line or git GUI applications and not requiring the DLB itself.
Additionally, its communication with a virtual fieldwork tool built in Unreal Engine opens up the possibility to be used for any immersive or non-immersive software built in this game engine.

\subsection{User Study \& Sampling}\label{evaluation-methodology}
We aimed at finding how researchers make use of the DLB and whether they are affected by the setting in which they have used the VFT.
We conducted a between-groups study defining the independent variable as the difference of using the VFT either on a desktop PC or in the dome.
Participant behavior was investigated by analyzing usage data from the resulting repositories and by gathering feedback on the "perceived usefulness" (PU) and "perceived ease of use" (PEOU) of the DLB, using the Technology Acceptance Model (TAM) questionnaire \cite{davis1989perceived, lewis2019comparison}.
Additionally, we qualitatively investigated responses to open-ended questions regarding reproducibility in the geosciences, which will be fully explored in a follow-up publication.

\paragraph{\textbf{Sampling:}} We recruited 18 researchers from our institution, 11 identifying as female and 7 as male, with a median age of 31.
Nine participants were doctoral researchers, two research software engineers, six post-docs or senior scientists holding a PhD title, and one professor.
None had previously used either the DLB or the VFT, but they were regularly engaged with research using geospatial data in different software environments.
Study participants were assigned to either the "Dome" or "Monitor" group alternately (to always ensure an almost matching group size) and could enter a raffle for a gift voucher (20€, 15€, of 10€) as compensation.



\paragraph{\textbf{Setup:}} Participants in the Dome condition used a motion-tracked Xbox controller for VFT input and had a separate PC monitor, mouse and keyboard available for using the DLB (see \cref{fig:vft_dome_setup}).
In the Monitor group, participants used keyboard and mouse for both VFT and DLB input and had two PC monitors.
To ensure that the different spatial perception of an office and the dome would not skew participants' experience the desktop PC sessions were also conducted in the dome at a desk and with projectors turned off.

\paragraph{\textbf{Procedure:}} Participants were introduced to the VFT and DLB by a researcher in an interactive introductory phase lasting 30 minutes. 
Afterwards, participants were given 30 minutes to explore and measure the 3D models in the VFT and use the DLB to record their thoughts and investigation.
To motivate DLB use, they were given a task scenario where their notes would support collaboration, review, and reproducibility in a geological research project.
The detailed task instructions are included in \cref{app:task}.


\paragraph{\textbf{Questionnaire:}} After the task, participants were given a questionnaire with questions on age, gender, their area of research and their usage of research software, as well as a slightly modified version of the TAM questionnaire asking participants about their agreement with twelve statements on a seven-point Likert scale  \cite{davis1989perceived, lewis2019comparison}.
The modified TAM questionnaire and explanation on our modification is included in \cref{app:tam}. 
Following the TAM questionnaire participants responded to ten optional open-ended questions on the three topics \textit{"Reproducibility in scientific work \& visualization in general"}, \textit{"Participants' personal experience with reproducibility in scientific work"}, and \textit{"The Digital Lab Book \& its role for reproducibility"}.


\section{Analysis \& Results}
\label{sec:results}


\subsection{Characterization of Participants}

\begin{figure}[!htbp]
    \centering
    \includegraphics[width=\linewidth]{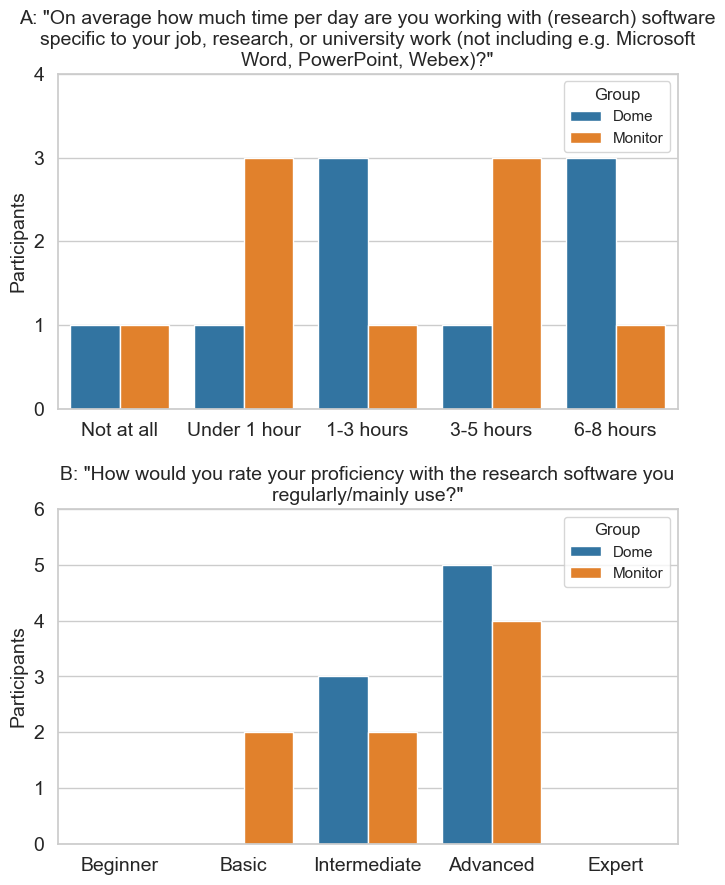}
    \caption{Self-reported use of research software per day and self-assessed proficiency with research software by participants. Visible is an overall slightly higher daily use of research software as well as slightly higher self-assessed proficiency in the Dome group.}
    \label{fig:software-usage}
    \Description{Self-reported use of research software per day and self-assessed proficiency with research software by participants. Visible is an overall slightly higher daily use of research software as well as slightly higher self-assessed proficiency in the Dome group.}
\end{figure}

We asked participants to indicate how frequently they are utilizing software for their research and how they would rate their expertise.
These self-assessments (see \cref{fig:software-usage}) indicate slightly higher frequency of research software usage in the "Dome" group and slightly higher self-assessed proficiency.

\subsection{Analysis of TAM Questionnaire}

\begin{figure}[!htbp]
\centering
    \centering
    \includegraphics[width=\linewidth]{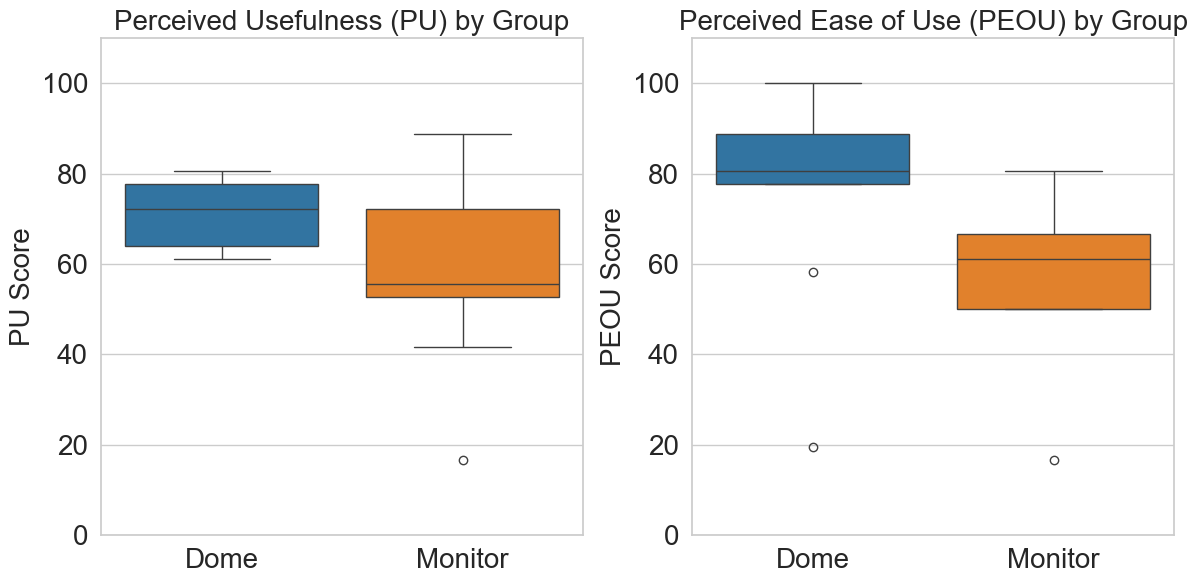}
    \caption{Perceived usefulness and perceived ease of use by participant group}
    \label{fig:pu-peou}
    \Description{Box plots of the perceived usefulness and perceived ease of use by participant group. The perceived usefulness on the left is visibly similar between groups. The perceived ease of use on the right shows stronger visible differences in favor of the Dome condition}

\end{figure}

Interested in knowing how participants would accept the DLB as a useful tool, we calculated PU and PEOU according to \citet{lewis2019comparison}.
The visible differences between groups in the PEOU (see \cref{fig:pu-peou}) warranted a closer look using statistical tests. Our null hypotheses were therefore

\begin{quote}
\textbf{H0\textsubscript{PU}}: There is no difference in the PU between the Dome and the Monitor group

\textbf{H0\textsubscript{PEOU}}: There is no difference in the PEOU between the Dome and the Monitor group
\end{quote}

with the alternative hypotheses that there is a difference introduced by the independent variable of the condition.

We used both the independent samples t-test (parametric) and the Mann-Whitney U test (non-parametric). The t-test assumptions were met, as Dome and Monitor groups were independent with no participant overlap. To address potential non-normality from outliers, we also performed a Mann-Whitney U test.
We utilize the implementations available through the Python library SciPy\cite{virtanen2020scipy}.

Results for the PU showed no significant differences between groups.
For the PEOU, the Mann-Whitney U test showed a significant difference (U = 16.5, p = 0.037 $<$ 0.05), the t-test, however, did not (see \cref{tab:stat-tests}). 
This hints at a slight difference in the PEOU between groups (higher in the Dome setting), although this could potentially be traced back to the higher software usage and self-assigned expertise level (as shown in \cref{fig:software-usage}).

We examined whether there was a difference in PU and PEOU when splitting participants not by study setting but by Low ($<$ 1 hour, 1-3 hours) vs. High (3-5 hours, 6-8 hours) software usage and Low (Basic, Intermediate) vs. High (Advanced, Expert) software proficiency. 
However, neither seemed to have any obvious impact on the PU nor PEOU as far as this can be assessed through hindsight groupings.

A figure displaying these results is included in the appendix (\cref{fig:tam_by_su}). Overall, both the PU and PEOU show that participants perceive the DLB generally positively with a median PU of 69.44 and a median PEOU of 72.22 (of a possible score of 100).
\begin{table}[ht]
    \caption{Comparison of PU and PEOU between Monitor and Dome groups}
    \label{tab:stat-tests}

    \begin{tabular*}{1\linewidth}{p{1.8cm}p{1.8cm}p{1.8cm}p{1.8cm}}
        \hline
        \textbf{Metric} & \textbf{Test} & \textbf{Statistic} & \textbf{p-value} \\
        \hline
        \multirow{2}{*}{PU} 
          & MWU & U = 25.0 & p = 0.183  \\
          & t-test & t = -1.59 & p = 0.141  \\
        \hline
        \multirow{2}{*}{PEOU} 
          & \textbf{\textit{MWU}} & \textit{\textbf{U = 16.5}} & \textit{\textbf{p = 0.037}}  \\
          & t-test & t = -1.72 & p = 0.106 \\
        \hline \\[-1em]
        \multicolumn{4}{l}{\parbox{\linewidth}{Median PU = \textbf{69.44},\\ Median PEOU = \textbf{72.22} (out of 100).}} \\
    \end{tabular*}
    \Description{A table showing the results of statistical tests using the Mann-Whitney-U and the independent samples t-test comparing the Perceived Usefulness and Perceived Ease of Use by participant group (MOnitor or Dome). Only the Mann-Whitney-U test showed a p-value below 0.05 with 0.037. }
\end{table}


\subsection{Quantitative Analysis of User Behavior}
\begin{figure}[!htbp]
    \centering
    \includegraphics[width=1\linewidth]{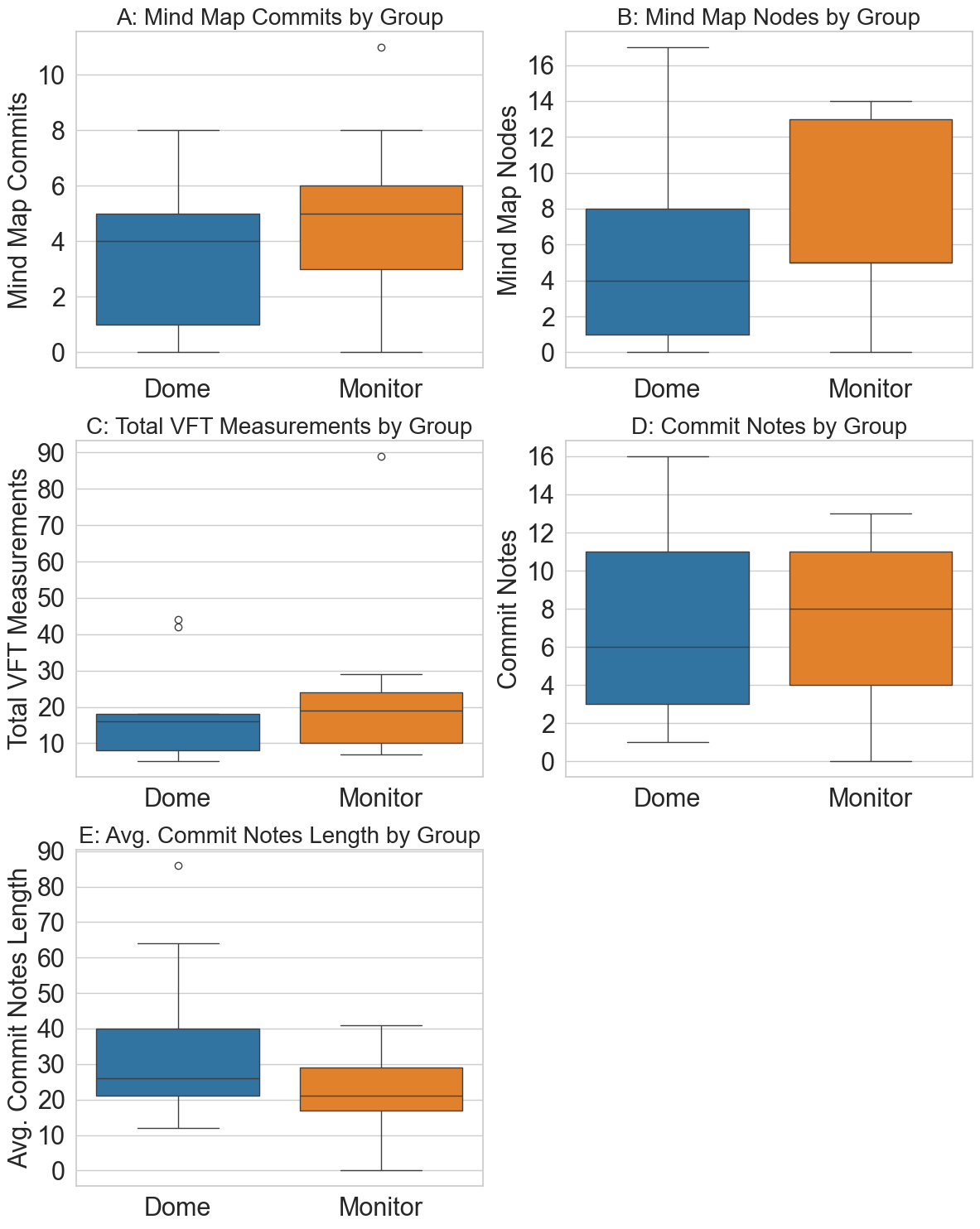}
    \caption{Participants' usage of the VFT and DLB by group. 
    \textbf{A:} How often participants interacted with the DLB Mind Map. The state of the Mind Map is saved as a commit every time the user navigates to another component again after modifying the Mind Map. 
    \textbf{B:} How much participants interacted with the DLB Mind Map, i.e. how many visualization states they added to the Mind Map.
    \textbf{C:} The total number of interactions in the form of measurements (location marker, distance etc.) in the VFT. Each measurement resulted in a separately saved visualization state.
    \textbf{D:} How many visualization states participants annotated with additional comments. 
    \textbf{E:} How much content participants filled their comments on visualization states with (number of characters).}
    \label{fig:dlb-usage}
    \Description{Participants' usage of the VFT and DLB by group. 
    A: How often participants interacted with the DLB Mind Map. The state of the Mind Map is saved as a commit every time the user navigates to another component again after modifying the Mind Map. 
    B: How much participants interacted with the DLB Mind Map, i.e. how many visualization states they added to the Mind Map.
    C: The total number of interactions in the form of measurements (location marker, distance etc.) in the VFT. Each measurement resulted in a separately saved visualization state.
    D: How many visualization states participants annotated with additional comments. 
    E: How much content participants filled their comments on visualization states with.}
\end{figure}

We were also interested in finding out whether there would be different usage patterns for the DLB and VFT depending on the group condition. 
We investigated several usage metrics gathered from the provenance repositories. \Cref{fig:dlb-usage} shows a selection of these statistics. 

The analysis revealed no meaningful differences in usage between the Dome and Monitor group. 
Using both previously used statistical tests also did not yield statistically significant results, suggesting that usage patterns did not differ.
Generally, participants used both the DLB and the VFT less than we anticipated, e.g. placing relatively few measurements or interacting little with the mind map (with the exception of very few outliers).
We hope that the qualitative analysis of the open-ended questions will reveal possible causes of this.
Most likely, though, this is due to the short time participants had to get familiar with the software and the task requiring to think and act as a researcher with a possibly entirely different area of research.

\subsection{Responses to Open-Ended Questions}
\label{subsec:open}

Participants generally acknowledged gaps in scientific reproducibility both in their respective fields overall and also in their own workflows.
Several reported the missing availability of software and data as an issue, stating e.g.

\begin{quote}
\emph{I feel like all the information is stored in people's computers or brains, and even when it's on git, zenodo or on a paper, it is still a very simplified, polished version of it, which might not tell the whole story.}
\end{quote}

Participants also raised concerns regarding the knowledge provenance of insights presented in papers:

\begin{quote}
    \emph{The owner of the data is often the expert, [...] so he/she can rate the reliaibility of a certain data point and therefore also build an argument maybe more on a certain part of the data than on another part. This knowledge is very hard to transfer to another person [...]. }
\end{quote}

Ideas to counteract this included designing standard provenance and reproducibility procedures and maybe more importantly forcing scientists to use them, although in some fields the requirements for this would be:

\begin{quote}
    \emph{Same samples, same machines, standard protocols, same people. Not possible!}
\end{quote}

While many of those who work with code reported using git for version control and reproducibility, participants working with visual GIS identified the missing of provenance information of the interactive workflow as an issue (which is in line with what \citet{ziegler2023needfinding} reported) and considered the DLB as a potentially helpful tool:

\begin{quote}
    \emph{It would make looking at models way more organized. It would force me to always record the things I am doing, and this makes it much easier to publish everything that is needed to reproduce the study.}
\end{quote}

Generally, the mere presence of a provenance tracking tool seemed to have changed participants' investigation behaviour, with several stating that the DLB encouraged them to approach the task \emph{in a more strategic or organized manner}.

Feedback on overall usability revealed several UI and UX improvements but also the necessity of more time to get familiar with both the VFT and the DLB to be able to use both productively.

\section{Conclusion}
We set out to evaluate a provenance management software for an immersive tool for interactive geoscientific 3D data visualization. With 18 researchers from an ocean research institute we investigated whether the provenance tool would be used differently if researchers interacted with the visualization in an immersive dome or in front of two PC monitors.

Participants generally perceived the provenance tool itself as both quite useful and easy to use, showing the general value of the design and its concept.
Meaningful differences between the groups were not observed except for higher PEOU in the Dome condition.
Investigating possible causes of this difference such as the self-assessed proficiency with research software, did not reveal separate influence on the PU/PEOU, suggesting that the ease of use was perceived as higher by participants due to the immersive setting.
This could be due to the physical separation between DLB and VFT, enabling participants to clearly separate what can be done and how to interact with each software. 
Participants in the monitor group might have had a harder time swapping back and forth mentally between applications.
Investigating usage patterns revealed no clear differences between groups, suggesting no notable correlation between group setting and tool usage.

Overall, the DLB was approached and judged positively by participants regardless whether it was used as companion to an immersive application or at a desk, implicating usefulness for both traditional research workflows in an office as well as in immersive environments.
This was also highlighted by participants' answers to open questions regarding reproducibilty in general and the use of the DLB.

The study has some limitations which should be adressed in follow-up research. 
Our sample size was affected by our qualitative research interest requiring participants being active in geoscientific research. 
A larger and more diverse sample size may expose further insights. Also, additional metrics such as task performance, accuracy or task completion were ommitted because of the exploratory nature of geological field work and our main interest in evaluating the DLB and VFT. 
The same applies to comparisons with
traditional documentation methods, for example comparing the NASA Task Load Index between groups using the DLB and groups using (hand)written notes.

\section*{Declaration of Generative AI and AI-assisted technologies in the writing process}
During the preparation of this work the authors used ChatGPT in order to check grammar and punctuation, find synonyms, and rephrase sentences to avoid repetitive text or redundant formulations. The authors reviewed and edited the content as needed and take full responsibility for the content of the publication. All content, interpretations, conclusions, and any remaining errors are our own.

\section*{Code Availability}
\label{subsec:code}
The source code of both the VFT as well as the DLB is available via our institution's gitlab:
\\
\\
\url{https://git.geomar.de/digital-lab-book/digital-lab-book}
\\
\url{https://git.geomar.de/arena/unreal-development/virtualfieldwork}

\begin{acks}
Armin Bernstetter is a doctoral researcher funded through the Helmholtz School for Marine Data Science (MarDATA), Grant No. HIDSS-0005.
\end{acks}
\newpage

\bibliographystyle{ACM-Reference-Format}
\bibliography{references}


\begin{thebibliography}{37}


\ifx \showCODEN    \undefined \def \showCODEN     #1{\unskip}     \fi
\ifx \showISBNx    \undefined \def \showISBNx     #1{\unskip}     \fi
\ifx \showISBNxiii \undefined \def \showISBNxiii  #1{\unskip}     \fi
\ifx \showISSN     \undefined \def \showISSN      #1{\unskip}     \fi
\ifx \showLCCN     \undefined \def \showLCCN      #1{\unskip}     \fi
\ifx \shownote     \undefined \def \shownote      #1{#1}          \fi
\ifx \showarticletitle \undefined \def \showarticletitle #1{#1}   \fi
\ifx \showURL      \undefined \def \showURL       {\relax}        \fi
\providecommand\bibfield[2]{#2}
\providecommand\bibinfo[2]{#2}
\providecommand\natexlab[1]{#1}
\providecommand\showeprint[2][]{arXiv:#2}

\bibitem[201(2013)]%
        {2013provoverview}
 \bibinfo{year}{2013}\natexlab{}.
\newblock \bibinfo{booktitle}{\emph{PROV-Overview. An Overview of the PROV
  Family of Documents}}.
\newblock \bibinfo{type}{Project Report}.
\newblock
\urldef\tempurl%
\url{https://eprints.soton.ac.uk/356854/}
\showURL{%
\tempurl}


\bibitem[Anderson et~al\mbox{.}(2006)]%
        {anderson2006visualization}
\bibfield{author}{\bibinfo{person}{Erik~W. Anderson},
  \bibinfo{person}{Steven~P. Callahan}, \bibinfo{person}{George T.~Y. Chen},
  \bibinfo{person}{Juliana Freire}, \bibinfo{person}{Emanuele Santos},
  \bibinfo{person}{Carlos~E. Scheidegger}, \bibinfo{person}{Claudio~T. Silva},
  {and} \bibinfo{person}{Huy~T. Vo}.} \bibinfo{year}{2006}\natexlab{}.
\newblock \bibinfo{booktitle}{\emph{Visualization in Radiation Oncology:
  Towards Replacing the Laboratory Notebook}}.
\newblock \bibinfo{type}{Technical Report} UUSCI-2006-17.
\newblock
\urldef\tempurl%
\url{http://www.sci.utah.edu/publications/SCITechReports/UUSCI-2006-017.pdf}
\showURL{%
\tempurl}


\bibitem[Arnaubec et~al\mbox{.}(2023)]%
        {arnaubec2023underwater}
\bibfield{author}{\bibinfo{person}{Aurélien Arnaubec}, \bibinfo{person}{Maxime
  Ferrera}, \bibinfo{person}{Javier Escartín}, \bibinfo{person}{Marjolaine
  Matabos}, \bibinfo{person}{Nuno Gracias}, {and} \bibinfo{person}{Jan
  Opderbecke}.} \bibinfo{year}{2023}\natexlab{}.
\newblock \showarticletitle{Underwater 3D Reconstruction from Video or Still
  Imagery: Matisse and 3DMetrics Processing and Exploitation Software}.
\newblock \bibinfo{journal}{\emph{Journal of Marine Science and Engineering}}
  \bibinfo{volume}{11}, \bibinfo{number}{5} (\bibinfo{year}{2023}).
\newblock
\showISSN{2077-1312}
\href{https://doi.org/10.3390/jmse11050985}{doi:\nolinkurl{10.3390/jmse11050985}}


\bibitem[Bernstetter et~al\mbox{.}(2025)]%
        {bernstetter2025virtualfieldwork}
\bibfield{author}{\bibinfo{person}{Armin Bernstetter}, \bibinfo{person}{Tom
  Kwasnitschka}, \bibinfo{person}{Jens Karstens}, \bibinfo{person}{Markus
  Schlüter}, {and} \bibinfo{person}{Isabella Peters}.}
  \bibinfo{year}{2025}\natexlab{}.
\newblock \showarticletitle{Virtual fieldwork in immersive environments using
  game engines}.
\newblock \bibinfo{journal}{\emph{Computers \& Geosciences}}
  \bibinfo{volume}{196} (\bibinfo{year}{2025}), \bibinfo{pages}{105855}.
\newblock
\showISSN{0098-3004}
\href{https://doi.org/10.1016/j.cageo.2025.105855}{doi:\nolinkurl{10.1016/j.cageo.2025.105855}}


\bibitem[Bernstetter et~al\mbox{.}(2023)]%
        {bernstetter2023practical}
\bibfield{author}{\bibinfo{person}{Armin Bernstetter}, \bibinfo{person}{Tom
  Kwasnitschka}, {and} \bibinfo{person}{Isabella Peters}.}
  \bibinfo{year}{2023}\natexlab{}.
\newblock \showarticletitle{{A Practical Approach to Provenance Capturing for
  Reproducible Visual Analytics at an Ocean Research Institute}}. In
  \bibinfo{booktitle}{\emph{EuroVis Workshop on Visual Analytics (EuroVA)}},
  \bibfield{editor}{\bibinfo{person}{Marco Angelini} {and}
  \bibinfo{person}{Mennatallah El-Assady}} (Eds.). \bibinfo{publisher}{The
  Eurographics Association}.
\newblock
\showISBNx{978-3-03868-222-6}
\showISSN{2664-4487}
\href{https://doi.org/10.2312/eurova.20231095}{doi:\nolinkurl{10.2312/eurova.20231095}}


\bibitem[Bonali et~al\mbox{.}(2024)]%
        {bonali2024geavr}
\bibfield{author}{\bibinfo{person}{Fabio~Luca Bonali}, \bibinfo{person}{Fabio
  Vitello}, \bibinfo{person}{Martin Kearl}, \bibinfo{person}{Alessandro
  Tibaldi}, \bibinfo{person}{Malcolm Whitworth}, \bibinfo{person}{Varvara
  Antoniou}, \bibinfo{person}{Elena Russo}, \bibinfo{person}{Emmanuel Delage},
  \bibinfo{person}{Paraskevi Nomikou}, \bibinfo{person}{Ugo Becciani},
  \bibinfo{person}{Benjamin {van Wyk de Vries}}, {and} \bibinfo{person}{Mel
  Krokos}.} \bibinfo{year}{2024}\natexlab{}.
\newblock \showarticletitle{GeaVR: An open-source tools package for
  geological-structural exploration and data collection using immersive virtual
  reality}.
\newblock \bibinfo{journal}{\emph{Applied Computing and Geosciences}}
  \bibinfo{volume}{21} (\bibinfo{year}{2024}), \bibinfo{pages}{100156}.
\newblock
\showISSN{2590-1974}
\href{https://doi.org/10.1016/j.acags.2024.100156}{doi:\nolinkurl{10.1016/j.acags.2024.100156}}


\bibitem[Callahan et~al\mbox{.}(2006)]%
        {callahan2006vistrails}
\bibfield{author}{\bibinfo{person}{Steven~P. Callahan},
  \bibinfo{person}{Juliana Freire}, \bibinfo{person}{Emanuele Santos},
  \bibinfo{person}{Carlos~E. Scheidegger}, \bibinfo{person}{Cl\'{a}udio~T.
  Silva}, {and} \bibinfo{person}{Huy~T. Vo}.} \bibinfo{year}{2006}\natexlab{}.
\newblock \showarticletitle{VisTrails: Visualization Meets Data Management}. In
  \bibinfo{booktitle}{\emph{Proceedings of the 2006 ACM SIGMOD International
  Conference on Management of Data}} (Chicago, IL, USA)
  \emph{(\bibinfo{series}{SIGMOD '06})}. \bibinfo{publisher}{Association for
  Computing Machinery}, \bibinfo{address}{New York, NY, USA},
  \bibinfo{pages}{745–747}.
\newblock
\showISBNx{1595934340}
\href{https://doi.org/10.1145/1142473.1142574}{doi:\nolinkurl{10.1145/1142473.1142574}}


\bibitem[Caravaca et~al\mbox{.}(2020)]%
        {caravaca2020digital}
\bibfield{author}{\bibinfo{person}{Gwénaël Caravaca},
  \bibinfo{person}{Stéphane {Le Mouélic}}, \bibinfo{person}{Nicolas Mangold},
  \bibinfo{person}{Jonas L’Haridon}, \bibinfo{person}{Laetitia {Le Deit}},
  {and} \bibinfo{person}{Marion Massé}.} \bibinfo{year}{2020}\natexlab{}.
\newblock \showarticletitle{3D digital outcrop model reconstruction of the
  Kimberley outcrop (Gale crater, Mars) and its integration into Virtual
  Reality for simulated geological analysis}.
\newblock \bibinfo{journal}{\emph{Planetary and Space Science}}
  \bibinfo{volume}{182} (\bibinfo{year}{2020}), \bibinfo{pages}{104808}.
\newblock
\showISSN{0032-0633}
\href{https://doi.org/10.1016/j.pss.2019.104808}{doi:\nolinkurl{10.1016/j.pss.2019.104808}}


\bibitem[Davis(1989)]%
        {davis1989perceived}
\bibfield{author}{\bibinfo{person}{Fred~D. Davis}.}
  \bibinfo{year}{1989}\natexlab{}.
\newblock \showarticletitle{Perceived Usefulness, Perceived Ease of Use, and
  User Acceptance of Information Technology}.
\newblock \bibinfo{journal}{\emph{MIS Quarterly}} \bibinfo{volume}{13},
  \bibinfo{number}{3} (\bibinfo{year}{1989}), \bibinfo{pages}{319--340}.
\newblock
\showISSN{02767783, 21629730}
\urldef\tempurl%
\url{http://www.jstor.org/stable/249008}
\showURL{%
\tempurl}


\bibitem[Del~Rio and da~Silva(2007a)]%
        {delrio2007identifying}
\bibfield{author}{\bibinfo{person}{Nicholas Del~Rio} {and}
  \bibinfo{person}{P~Pinheiro da Silva}.} \bibinfo{year}{2007}\natexlab{a}.
\newblock \showarticletitle{Identifying and explaining map imperfections
  through knowledge provenance visualization}.
\newblock \bibinfo{journal}{\emph{University of Texas at El Paso}}
  (\bibinfo{year}{2007}).
\newblock


\bibitem[Del~Rio and da~Silva(2007b)]%
        {delrio2007probeit}
\bibfield{author}{\bibinfo{person}{Nicholas Del~Rio} {and}
  \bibinfo{person}{Paulo~Pinheiro da Silva}.} \bibinfo{year}{2007}\natexlab{b}.
\newblock \showarticletitle{Probe-It! Visualization Support for Provenance}. In
  \bibinfo{booktitle}{\emph{Advances in Visual Computing}},
  \bibfield{editor}{\bibinfo{person}{George Bebis}, \bibinfo{person}{Richard
  Boyle}, \bibinfo{person}{Bahram Parvin}, \bibinfo{person}{Darko Koracin},
  \bibinfo{person}{Nikos Paragios}, \bibinfo{person}{Syeda-Mahmood Tanveer},
  \bibinfo{person}{Tao Ju}, \bibinfo{person}{Zicheng Liu},
  \bibinfo{person}{Sabine Coquillart}, \bibinfo{person}{Carolina Cruz-Neira},
  \bibinfo{person}{Torsten M{\"u}ller}, {and} \bibinfo{person}{Tom Malzbender}}
  (Eds.). \bibinfo{publisher}{Springer Berlin Heidelberg},
  \bibinfo{address}{Berlin, Heidelberg}, \bibinfo{pages}{732--741}.
\newblock
\showISBNx{978-3-540-76856-2}


\bibitem[Fekete et~al\mbox{.}(2019)]%
        {fekete2019provenance}
\bibfield{author}{\bibinfo{person}{Jean-Daniel Fekete}, \bibinfo{person}{T.~J.
  Jankun-Kelly}, \bibinfo{person}{Melanie Tory}, {and} \bibinfo{person}{Kai
  Xu}.} \bibinfo{year}{2019}\natexlab{}.
\newblock \showarticletitle{Provenance and Logging for Sense Making (Dagstuhl
  Seminar 18462)}.
\newblock \bibinfo{journal}{\emph{Dagstuhl Reports}} \bibinfo{volume}{8},
  \bibinfo{number}{11} (\bibinfo{year}{2019}), \bibinfo{pages}{35--62}.
\newblock
\showISSN{2192-5283}
\href{https://doi.org/10.4230/DagRep.8.11.35}{doi:\nolinkurl{10.4230/DagRep.8.11.35}}


\bibitem[Fidler and Wilcox(2021)]%
        {fidler2021reproducibility}
\bibfield{author}{\bibinfo{person}{Fiona Fidler} {and} \bibinfo{person}{John
  Wilcox}.} \bibinfo{year}{2021}\natexlab{}.
\newblock \showarticletitle{{Reproducibility of Scientific Results}}.
\newblock In \bibinfo{booktitle}{\emph{The {Stanford} Encyclopedia of
  Philosophy} (\bibinfo{edition}{{S}ummer 2021} ed.)},
  \bibfield{editor}{\bibinfo{person}{Edward~N. Zalta}} (Ed.).
  \bibinfo{publisher}{Metaphysics Research Lab, Stanford University}.
\newblock
\urldef\tempurl%
\url{https://plato.stanford.edu/archives/sum2021/entries/scientific-reproducibility/}
\showURL{%
\tempurl}


\bibitem[Frew and Bose(2001)]%
        {frew2001earth}
\bibfield{author}{\bibinfo{person}{J. Frew} {and} \bibinfo{person}{R. Bose}.}
  \bibinfo{year}{2001}\natexlab{}.
\newblock \showarticletitle{Earth System Science Workbench: a data management
  infrastructure for earth science products}. In
  \bibinfo{booktitle}{\emph{Proceedings Thirteenth International Conference on
  Scientific and Statistical Database Management. SSDBM 2001}}.
  \bibinfo{pages}{180--189}.
\newblock
\showISSN{1099-3371}
\href{https://doi.org/10.1109/SSDM.2001.938550}{doi:\nolinkurl{10.1109/SSDM.2001.938550}}


\bibitem[Howe et~al\mbox{.}(2008)]%
        {howe2008endtoend}
\bibfield{author}{\bibinfo{person}{Bill Howe}, \bibinfo{person}{Peter Lawson},
  \bibinfo{person}{Renee Bellinger}, \bibinfo{person}{Erik Anderson},
  \bibinfo{person}{Emanuele Santos}, \bibinfo{person}{Juliana Freire},
  \bibinfo{person}{Carlos Scheidegger}, \bibinfo{person}{António Baptista},
  {and} \bibinfo{person}{Cláudio Silva}.} \bibinfo{year}{2008}\natexlab{}.
\newblock \showarticletitle{End-to-End eScience: Integrating Workflow, Query,
  Visualization, and Provenance at an Ocean Observatory}. In
  \bibinfo{booktitle}{\emph{2008 IEEE Fourth International Conference on
  eScience}}. \bibinfo{pages}{127--134}.
\newblock
\href{https://doi.org/10.1109/eScience.2008.67}{doi:\nolinkurl{10.1109/eScience.2008.67}}


\bibitem[Klippel et~al\mbox{.}(2019)]%
        {klippel2019transforming}
\bibfield{author}{\bibinfo{person}{Alexander Klippel}, \bibinfo{person}{Jiayan
  Zhao}, \bibinfo{person}{Kathy~Lou Jackson}, \bibinfo{person}{Peter~La
  Femina}, \bibinfo{person}{Chris Stubbs}, \bibinfo{person}{Ryan Wetzel},
  \bibinfo{person}{Jordan Blair}, \bibinfo{person}{Jan~Oliver Wallgrün}, {and}
  \bibinfo{person}{Danielle Oprean}.} \bibinfo{year}{2019}\natexlab{}.
\newblock \showarticletitle{Transforming Earth Science Education Through
  Immersive Experiences: Delivering on a Long Held Promise}.
\newblock \bibinfo{journal}{\emph{Journal of Educational Computing Research}}
  \bibinfo{volume}{57}, \bibinfo{number}{7} (\bibinfo{year}{2019}),
  \bibinfo{pages}{1745--1771}.
\newblock
\showeprint{https://doi.org/10.1177/0735633119854025}
\href{https://doi.org/10.1177/0735633119854025}{doi:\nolinkurl{10.1177/0735633119854025}}


\bibitem[Klippel et~al\mbox{.}(2020)]%
        {klippel2020value}
\bibfield{author}{\bibinfo{person}{Alexander Klippel}, \bibinfo{person}{Jiayan
  Zhao}, \bibinfo{person}{Danielle Oprean}, \bibinfo{person}{Jan~Oliver
  Wallgr{\"u}n}, \bibinfo{person}{Chris Stubbs}, \bibinfo{person}{Peter
  La~Femina}, {and} \bibinfo{person}{Kathy~L. Jackson}.}
  \bibinfo{year}{2020}\natexlab{}.
\newblock \showarticletitle{The value of being there: toward a science of
  immersive virtual field trips}.
\newblock \bibinfo{journal}{\emph{Virtual Reality}} \bibinfo{volume}{24},
  \bibinfo{number}{4} (\bibinfo{date}{01 Dec.} \bibinfo{year}{2020}),
  \bibinfo{pages}{753--770}.
\newblock
\showISSN{1434-9957}
\href{https://doi.org/10.1007/s10055-019-00418-5}{doi:\nolinkurl{10.1007/s10055-019-00418-5}}


\bibitem[Kwasnitschka et~al\mbox{.}(2013)]%
        {kwasnitschka2013doing}
\bibfield{author}{\bibinfo{person}{Tom Kwasnitschka}, \bibinfo{person}{Thor~H.
  Hansteen}, \bibinfo{person}{Colin~W. Devey}, {and} \bibinfo{person}{Steffen
  Kutterolf}.} \bibinfo{year}{2013}\natexlab{}.
\newblock \showarticletitle{Doing fieldwork on the seafloor: Photogrammetric
  techniques to yield 3D visual models from ROV video}.
\newblock \bibinfo{journal}{\emph{Computers \& Geosciences}}
  \bibinfo{volume}{52} (\bibinfo{year}{2013}), \bibinfo{pages}{218--226}.
\newblock
\showISSN{0098-3004}
\href{https://doi.org/10.1016/j.cageo.2012.10.008}{doi:\nolinkurl{10.1016/j.cageo.2012.10.008}}


\bibitem[Kwasnitschka et~al\mbox{.}(2023)]%
        {kwasnitschka2023spatially}
\bibfield{author}{\bibinfo{person}{Tom Kwasnitschka}, \bibinfo{person}{Markus
  Schlüter}, \bibinfo{person}{Jens Klimmeck}, \bibinfo{person}{Armin
  Bernstetter}, \bibinfo{person}{Felix Gross}, {and} \bibinfo{person}{Isabella
  Peters}.} \bibinfo{year}{2023}\natexlab{}.
\newblock \showarticletitle{Spatially Immersive Visualization Domes as a Marine
  Geoscientific Research Tool}. In \bibinfo{booktitle}{\emph{Workshop on
  Visualisation in Environmental Sciences (EnvirVis)}},
  \bibfield{editor}{\bibinfo{person}{Soumya Dutta}, \bibinfo{person}{Kathrin
  Feige}, \bibinfo{person}{Karsten Rink}, {and} \bibinfo{person}{Dirk Zeckzer}}
  (Eds.). \bibinfo{publisher}{The Eurographics Association}.
\newblock
\showISBNx{978-3-03868-223-3}
\href{https://doi.org/10.2312/envirvis.20231102}{doi:\nolinkurl{10.2312/envirvis.20231102}}


\bibitem[Lewis(2019)]%
        {lewis2019comparison}
\bibfield{author}{\bibinfo{person}{James~R. Lewis}.}
  \bibinfo{year}{2019}\natexlab{}.
\newblock \showarticletitle{Comparison of four TAM item formats: effect of
  response option labels and order}.
\newblock \bibinfo{journal}{\emph{J. Usability Studies}} \bibinfo{volume}{14},
  \bibinfo{number}{4} (\bibinfo{date}{1 8} \bibinfo{year}{2019}),
  \bibinfo{pages}{224–236}.
\newblock


\bibitem[Liu et~al\mbox{.}(2019)]%
        {liu2019improving}
\bibfield{author}{\bibinfo{person}{Zhong Liu}, \bibinfo{person}{Jianwu Wang},
  \bibinfo{person}{Shimei Pan}, {and} \bibinfo{person}{David~J Meyer}.}
  \bibinfo{year}{2019}\natexlab{}.
\newblock \showarticletitle{Improving reproducibility in Earth science
  Research}.
\newblock \bibinfo{journal}{\emph{Eos}} \bibinfo{number}{100}
  (\bibinfo{date}{Oct.} \bibinfo{year}{2019}).
\newblock
\urldef\tempurl%
\url{https://doi.org/10.1029/2019EO136216}
\showURL{%
\tempurl}


\bibitem[Marriott et~al\mbox{.}(2018)]%
        {marriott2018immersive}
\bibfield{author}{\bibinfo{person}{Kim Marriott}, \bibinfo{person}{Falk
  Schreiber}, \bibinfo{person}{Tim Dwyer}, \bibinfo{person}{Karsten Klein},
  \bibinfo{person}{Nathalie~Henry Riche}, \bibinfo{person}{Takayuki Itoh},
  \bibinfo{person}{Wolfgang Stuerzlinger}, {and} \bibinfo{person}{Bruce~H
  Thomas}.} \bibinfo{year}{2018}\natexlab{}.
\newblock \bibinfo{booktitle}{\emph{Immersive analytics}}.
  Vol.~\bibinfo{volume}{11190}.
\newblock \bibinfo{publisher}{Springer}.
\newblock


\bibitem[M\'etois et~al\mbox{.}(2021)]%
        {metois2021oceanic}
\bibfield{author}{\bibinfo{person}{M. M\'etois}, \bibinfo{person}{J.-E.
  Martelat}, \bibinfo{person}{J. Billant}, \bibinfo{person}{M. Andreani},
  \bibinfo{person}{J. Escart\'{\i}n}, \bibinfo{person}{F. Leclerc}, {and}
  \bibinfo{person}{the ICAP~team}.} \bibinfo{year}{2021}\natexlab{}.
\newblock \showarticletitle{Deep oceanic submarine fieldwork with undergraduate
  students: an immersive experience with the Minerve software}.
\newblock \bibinfo{journal}{\emph{Solid Earth}} \bibinfo{volume}{12},
  \bibinfo{number}{12} (\bibinfo{year}{2021}), \bibinfo{pages}{2789--2802}.
\newblock
\href{https://doi.org/10.5194/se-12-2789-2021}{doi:\nolinkurl{10.5194/se-12-2789-2021}}


\bibitem[Moreau et~al\mbox{.}(2011)]%
        {moreau2011provenance}
\bibfield{author}{\bibinfo{person}{Luc Moreau}, \bibinfo{person}{Ben Clifford},
  \bibinfo{person}{Juliana Freire}, \bibinfo{person}{Joe Futrelle},
  \bibinfo{person}{Yolanda Gil}, \bibinfo{person}{Paul Groth},
  \bibinfo{person}{Natalia Kwasnikowska}, \bibinfo{person}{Simon Miles},
  \bibinfo{person}{Paolo Missier}, \bibinfo{person}{Jim Myers},
  \bibinfo{person}{Beth Plale}, \bibinfo{person}{Yogesh Simmhan},
  \bibinfo{person}{Eric Stephan}, {and} \bibinfo{person}{Jan~Van {den
  Bussche}}.} \bibinfo{year}{2011}\natexlab{}.
\newblock \showarticletitle{The Open Provenance Model core specification
  (v1.1)}.
\newblock \bibinfo{journal}{\emph{Future Generation Computer Systems}}
  \bibinfo{volume}{27}, \bibinfo{number}{6} (\bibinfo{year}{2011}),
  \bibinfo{pages}{743--756}.
\newblock
\showISSN{0167-739X}
\href{https://doi.org/10.1016/j.future.2010.07.005}{doi:\nolinkurl{10.1016/j.future.2010.07.005}}


\bibitem[Pirolli and Card(2005)]%
        {pirolli2005sensemaking}
\bibfield{author}{\bibinfo{person}{Peter Pirolli} {and} \bibinfo{person}{Stuart
  Card}.} \bibinfo{year}{2005}\natexlab{}.
\newblock \showarticletitle{The sensemaking process and leverage points for
  analyst technology as identified through cognitive task analysis}. In
  \bibinfo{booktitle}{\emph{Proceedings of international conference on
  intelligence analysis}}, Vol.~\bibinfo{volume}{5}. McLean, VA, USA,
  \bibinfo{pages}{2--4}.
\newblock


\bibitem[Ragan et~al\mbox{.}(2015)]%
        {ragan2015evaluating}
\bibfield{author}{\bibinfo{person}{Eric~D. Ragan}, \bibinfo{person}{John~R.
  Goodall}, {and} \bibinfo{person}{Albert Tung}.}
  \bibinfo{year}{2015}\natexlab{}.
\newblock \showarticletitle{Evaluating How Level of Detail of Visual History
  Affects Process Memory}. In \bibinfo{booktitle}{\emph{Proceedings of the 33rd
  Annual ACM Conference on Human Factors in Computing Systems}} (Seoul,
  Republic of Korea) \emph{(\bibinfo{series}{CHI '15})}.
  \bibinfo{publisher}{Association for Computing Machinery},
  \bibinfo{address}{New York, NY, USA}, \bibinfo{pages}{2711–2720}.
\newblock
\showISBNx{9781450331456}
\href{https://doi.org/10.1145/2702123.2702376}{doi:\nolinkurl{10.1145/2702123.2702376}}


\bibitem[Schuchardt and Bowman(2007)]%
        {schuchardt2007benefits}
\bibfield{author}{\bibinfo{person}{Philip Schuchardt} {and}
  \bibinfo{person}{Doug~A. Bowman}.} \bibinfo{year}{2007}\natexlab{}.
\newblock \showarticletitle{The benefits of immersion for spatial understanding
  of complex underground cave systems}. In
  \bibinfo{booktitle}{\emph{Proceedings of the 2007 {ACM} symposium on Virtual
  reality software and technology - {VRST} {\textquotesingle}07}}.
  \bibinfo{publisher}{{ACM} Press}.
\newblock
\href{https://doi.org/10.1145/1315184.1315205}{doi:\nolinkurl{10.1145/1315184.1315205}}


\bibitem[Shrinivasan and van Wijk(2008)]%
        {shrinivasan2008supporting}
\bibfield{author}{\bibinfo{person}{Yedendra~Babu Shrinivasan} {and}
  \bibinfo{person}{Jarke~J. van Wijk}.} \bibinfo{year}{2008}\natexlab{}.
\newblock \showarticletitle{Supporting the analytical reasoning process in
  information visualization}. In \bibinfo{booktitle}{\emph{Proceedings of the
  {SIGCHI} Conference on Human Factors in Computing Systems}}.
  \bibinfo{publisher}{{ACM}}.
\newblock
\href{https://doi.org/10.1145/1357054.1357247}{doi:\nolinkurl{10.1145/1357054.1357247}}


\bibitem[Steventon et~al\mbox{.}(2022)]%
        {steventon2022reproducibility}
\bibfield{author}{\bibinfo{person}{Michael~J. Steventon},
  \bibinfo{person}{Christopher A-L. Jackson}, \bibinfo{person}{Matt Hall},
  \bibinfo{person}{Mark~T. Ireland}, \bibinfo{person}{Marcus Munafo}, {and}
  \bibinfo{person}{Kathryn~J. Roberts}.} \bibinfo{year}{2022}\natexlab{}.
\newblock \showarticletitle{Reproducibility in Subsurface Geoscience}.
\newblock \bibinfo{journal}{\emph{Earth Science, Systems and Society}}
  \bibinfo{volume}{2}, \bibinfo{number}{1} (\bibinfo{year}{2022}),
  \bibinfo{pages}{10051}.
\newblock
\showeprint{https://www.lyellcollection.org/doi/pdf/10.3389/esss.2022.10051}
\href{https://doi.org/10.3389/esss.2022.10051}{doi:\nolinkurl{10.3389/esss.2022.10051}}


\bibitem[Toniolo et~al\mbox{.}(2015)]%
        {toniolo2015supporting}
\bibfield{author}{\bibinfo{person}{Alice Toniolo}, \bibinfo{person}{Timothy~J.
  Norman}, \bibinfo{person}{Anthony Etuk}, \bibinfo{person}{Federico Cerutti},
  \bibinfo{person}{Robin~Wentao Ouyang}, \bibinfo{person}{Mani Srivastava},
  \bibinfo{person}{Nir Oren}, \bibinfo{person}{Timothy Dropps},
  \bibinfo{person}{John~A. Allen}, {and} \bibinfo{person}{Paul Sullivan}.}
  \bibinfo{year}{2015}\natexlab{}.
\newblock \showarticletitle{Supporting Reasoning with Different Types of
  Evidence in Intelligence Analysis}. In \bibinfo{booktitle}{\emph{Proceedings
  of the 2015 International Conference on Autonomous Agents and Multiagent
  Systems}} (Istanbul, Turkey) \emph{(\bibinfo{series}{AAMAS '15})}.
  \bibinfo{publisher}{International Foundation for Autonomous Agents and
  Multiagent Systems}, \bibinfo{address}{Richland, SC},
  \bibinfo{pages}{781â€“789}.
\newblock
\showISBNx{9781450334136}


\bibitem[van Wijk(2005)]%
        {vanwijk2005value}
\bibfield{author}{\bibinfo{person}{Jarke~J. van Wijk}.}
  \bibinfo{year}{2005}\natexlab{}.
\newblock \showarticletitle{The Value of Visualization.}. In
  \bibinfo{booktitle}{\emph{IEEE Visualization}}. \bibinfo{publisher}{IEEE
  Computer Society}, \bibinfo{pages}{79--86}.
\newblock
\showISBNx{0-7803-9462-3}
\href{https://doi.org/10.1109/VISUAL.2005.1532781}{doi:\nolinkurl{10.1109/VISUAL.2005.1532781}}


\bibitem[Virtanen et~al\mbox{.}(2020)]%
        {virtanen2020scipy}
\bibfield{author}{\bibinfo{person}{Pauli Virtanen}, \bibinfo{person}{Ralf
  Gommers}, \bibinfo{person}{Travis~E. Oliphant}, \bibinfo{person}{Matt
  Haberland}, \bibinfo{person}{Tyler Reddy}, \bibinfo{person}{David
  Cournapeau}, \bibinfo{person}{Evgeni Burovski}, \bibinfo{person}{Pearu
  Peterson}, \bibinfo{person}{Warren Weckesser}, \bibinfo{person}{Jonathan
  Bright}, \bibinfo{person}{Stéfan~J. van~der Walt}, \bibinfo{person}{Matthew
  Brett}, \bibinfo{person}{Joshua Wilson}, \bibinfo{person}{K.~Jarrod Millman},
  \bibinfo{person}{Nikolay Mayorov}, \bibinfo{person}{Andrew R.~J. Nelson},
  \bibinfo{person}{Eric Jones}, \bibinfo{person}{Robert Kern},
  \bibinfo{person}{Eric Larson}, \bibinfo{person}{C~J Carey},
  \bibinfo{person}{İlhan Polat}, \bibinfo{person}{Yu Feng},
  \bibinfo{person}{Eric~W. Moore}, \bibinfo{person}{Jake VanderPlas},
  \bibinfo{person}{Denis Laxalde}, \bibinfo{person}{Josef Perktold},
  \bibinfo{person}{Robert Cimrman}, \bibinfo{person}{Ian Henriksen},
  \bibinfo{person}{E.~A. Quintero}, \bibinfo{person}{Charles~R. Harris},
  \bibinfo{person}{Anne~M. Archibald}, \bibinfo{person}{Antônio~H. Ribeiro},
  \bibinfo{person}{Fabian Pedregosa}, \bibinfo{person}{Paul van Mulbregt},
  \bibinfo{person}{Aditya Vijaykumar}, \bibinfo{person}{Alessandro~Pietro
  Bardelli}, \bibinfo{person}{Alex Rothberg}, \bibinfo{person}{Andreas
  Hilboll}, \bibinfo{person}{Andreas Kloeckner}, \bibinfo{person}{Anthony
  Scopatz}, \bibinfo{person}{Antony Lee}, \bibinfo{person}{Ariel Rokem},
  \bibinfo{person}{C.~Nathan Woods}, \bibinfo{person}{Chad Fulton},
  \bibinfo{person}{Charles Masson}, \bibinfo{person}{Christian Häggström},
  \bibinfo{person}{Clark Fitzgerald}, \bibinfo{person}{David~A. Nicholson},
  \bibinfo{person}{David~R. Hagen}, \bibinfo{person}{Dmitrii~V. Pasechnik},
  \bibinfo{person}{Emanuele Olivetti}, \bibinfo{person}{Eric Martin},
  \bibinfo{person}{Eric Wieser}, \bibinfo{person}{Fabrice Silva},
  \bibinfo{person}{Felix Lenders}, \bibinfo{person}{Florian Wilhelm},
  \bibinfo{person}{G. Young}, \bibinfo{person}{Gavin~A. Price},
  \bibinfo{person}{Gert-Ludwig Ingold}, \bibinfo{person}{Gregory~E. Allen},
  \bibinfo{person}{Gregory~R. Lee}, \bibinfo{person}{Hervé Audren},
  \bibinfo{person}{Irvin Probst}, \bibinfo{person}{Jörg~P. Dietrich},
  \bibinfo{person}{Jacob Silterra}, \bibinfo{person}{James~T Webber},
  \bibinfo{person}{Janko Slavič}, \bibinfo{person}{Joel Nothman},
  \bibinfo{person}{Johannes Buchner}, \bibinfo{person}{Johannes Kulick},
  \bibinfo{person}{Johannes~L. Schönberger}, \bibinfo{person}{José~Vinícius
  de Miranda~Cardoso}, \bibinfo{person}{Joscha Reimer}, \bibinfo{person}{Joseph
  Harrington}, \bibinfo{person}{Juan Luis~Cano Rodríguez},
  \bibinfo{person}{Juan Nunez-Iglesias}, \bibinfo{person}{Justin Kuczynski},
  \bibinfo{person}{Kevin Tritz}, \bibinfo{person}{Martin Thoma},
  \bibinfo{person}{Matthew Newville}, \bibinfo{person}{Matthias Kümmerer},
  \bibinfo{person}{Maximilian Bolingbroke}, \bibinfo{person}{Michael Tartre},
  \bibinfo{person}{Mikhail Pak}, \bibinfo{person}{Nathaniel~J. Smith},
  \bibinfo{person}{Nikolai Nowaczyk}, \bibinfo{person}{Nikolay Shebanov},
  \bibinfo{person}{Oleksandr Pavlyk}, \bibinfo{person}{Per~A. Brodtkorb},
  \bibinfo{person}{Perry Lee}, \bibinfo{person}{Robert~T. McGibbon},
  \bibinfo{person}{Roman Feldbauer}, \bibinfo{person}{Sam Lewis},
  \bibinfo{person}{Sam Tygier}, \bibinfo{person}{Scott Sievert},
  \bibinfo{person}{Sebastiano Vigna}, \bibinfo{person}{Stefan Peterson},
  \bibinfo{person}{Surhud More}, \bibinfo{person}{Tadeusz Pudlik},
  \bibinfo{person}{Takuya Oshima}, \bibinfo{person}{Thomas~J. Pingel},
  \bibinfo{person}{Thomas~P. Robitaille}, \bibinfo{person}{Thomas Spura},
  \bibinfo{person}{Thouis~R. Jones}, \bibinfo{person}{Tim Cera},
  \bibinfo{person}{Tim Leslie}, \bibinfo{person}{Tiziano Zito},
  \bibinfo{person}{Tom Krauss}, \bibinfo{person}{Utkarsh Upadhyay},
  \bibinfo{person}{Yaroslav~O. Halchenko}, {and} \bibinfo{person}{Yoshiki
  Vázquez-Baeza}.} \bibinfo{year}{2020}\natexlab{}.
\newblock \showarticletitle{SciPy 1.0: fundamental algorithms for scientific
  computing in Python}.
\newblock \bibinfo{journal}{\emph{Nature Methods}} \bibinfo{volume}{17},
  \bibinfo{number}{3} (\bibinfo{date}{Feb.} \bibinfo{year}{2020}),
  \bibinfo{pages}{261–272}.
\newblock
\showISSN{1548-7105}
\href{https://doi.org/10.1038/s41592-019-0686-2}{doi:\nolinkurl{10.1038/s41592-019-0686-2}}


\bibitem[Wexelblat and Maes(1999)]%
        {wexelblat1999footprints}
\bibfield{author}{\bibinfo{person}{Alan Wexelblat} {and}
  \bibinfo{person}{Pattie Maes}.} \bibinfo{year}{1999}\natexlab{}.
\newblock \showarticletitle{Footprints: History-Rich Tools for Information
  Foraging}. In \bibinfo{booktitle}{\emph{Proceedings of the SIGCHI Conference
  on Human Factors in Computing Systems}} (Pittsburgh, Pennsylvania, USA)
  \emph{(\bibinfo{series}{CHI '99})}. \bibinfo{publisher}{Association for
  Computing Machinery}, \bibinfo{address}{New York, NY, USA},
  \bibinfo{pages}{270–277}.
\newblock
\showISBNx{0201485591}
\href{https://doi.org/10.1145/302979.303060}{doi:\nolinkurl{10.1145/302979.303060}}


\bibitem[Xu et~al\mbox{.}(2020)]%
        {xu2020survey}
\bibfield{author}{\bibinfo{person}{Kai Xu}, \bibinfo{person}{Alvitta Ottley},
  \bibinfo{person}{Conny Walchshofer}, \bibinfo{person}{Marc Streit},
  \bibinfo{person}{Remco Chang}, {and} \bibinfo{person}{John Wenskovitch}.}
  \bibinfo{year}{2020}\natexlab{}.
\newblock \showarticletitle{Survey on the Analysis of User Interactions and
  Visualization Provenance}.
\newblock \bibinfo{journal}{\emph{Computer Graphics Forum}}
  \bibinfo{volume}{39}, \bibinfo{number}{3} (\bibinfo{year}{2020}),
  \bibinfo{pages}{757--783}.
\newblock
\showeprint{https://onlinelibrary.wiley.com/doi/pdf/10.1111/cgf.14035}
\href{https://doi.org/10.1111/cgf.14035}{doi:\nolinkurl{10.1111/cgf.14035}}


\bibitem[Zhang et~al\mbox{.}(2023)]%
        {zhang2023defining}
\bibfield{author}{\bibinfo{person}{Yidan Zhang}, \bibinfo{person}{Barrett Ens},
  \bibinfo{person}{Kadek~Ananta Satriadi}, \bibinfo{person}{Ying Yang}, {and}
  \bibinfo{person}{Sarah Goodwin}.} \bibinfo{year}{2023}\natexlab{}.
\newblock \showarticletitle{Defining Embodied Provenance for Immersive
  Sensemaking}. In \bibinfo{booktitle}{\emph{Extended Abstracts of the 2023
  {CHI} Conference on Human Factors in Computing Systems}}.
  \bibinfo{publisher}{{ACM}}.
\newblock
\href{https://doi.org/10.1145/3544549.3585691}{doi:\nolinkurl{10.1145/3544549.3585691}}


\bibitem[Zhao et~al\mbox{.}(2019)]%
        {zhao2019harnessing}
\bibfield{author}{\bibinfo{person}{Jiayan Zhao}, \bibinfo{person}{Jan~Oliver
  Wallgrün}, \bibinfo{person}{Peter~C. LaFemina}, \bibinfo{person}{Jim
  Normandeau}, {and} \bibinfo{person}{Alexander Klippel}.}
  \bibinfo{year}{2019}\natexlab{}.
\newblock \showarticletitle{Harnessing the power of immersive virtual reality -
  visualization and analysis of 3D earth science data sets}.
\newblock \bibinfo{journal}{\emph{Geo-spatial Information Science}}
  \bibinfo{volume}{22}, \bibinfo{number}{4} (\bibinfo{year}{2019}),
  \bibinfo{pages}{237--250}.
\newblock
\href{https://doi.org/10.1080/10095020.2019.1621544}{doi:\nolinkurl{10.1080/10095020.2019.1621544}}


\bibitem[Ziegler and Chasins(2023)]%
        {ziegler2023needfinding}
\bibfield{author}{\bibinfo{person}{Parker Ziegler} {and}
  \bibinfo{person}{Sarah~E. Chasins}.} \bibinfo{year}{2023}\natexlab{}.
\newblock \showarticletitle{A Need-Finding Study with Users of Geospatial
  Data}. In \bibinfo{booktitle}{\emph{Proceedings of the 2023 {CHI} Conference
  on Human Factors in Computing Systems}}. \bibinfo{publisher}{{ACM}}.
\newblock
\href{https://doi.org/10.1145/3544548.3581370}{doi:\nolinkurl{10.1145/3544548.3581370}}


\end{thebibliography}


\appendix
\renewcommand{\thefigure}{A.\arabic{figure}}
\setcounter{figure}{0}  

\section{Additional Figures}
\label{app:figs}


\begin{figure}[!htbp]
    \centering
    \includegraphics[width=\linewidth]{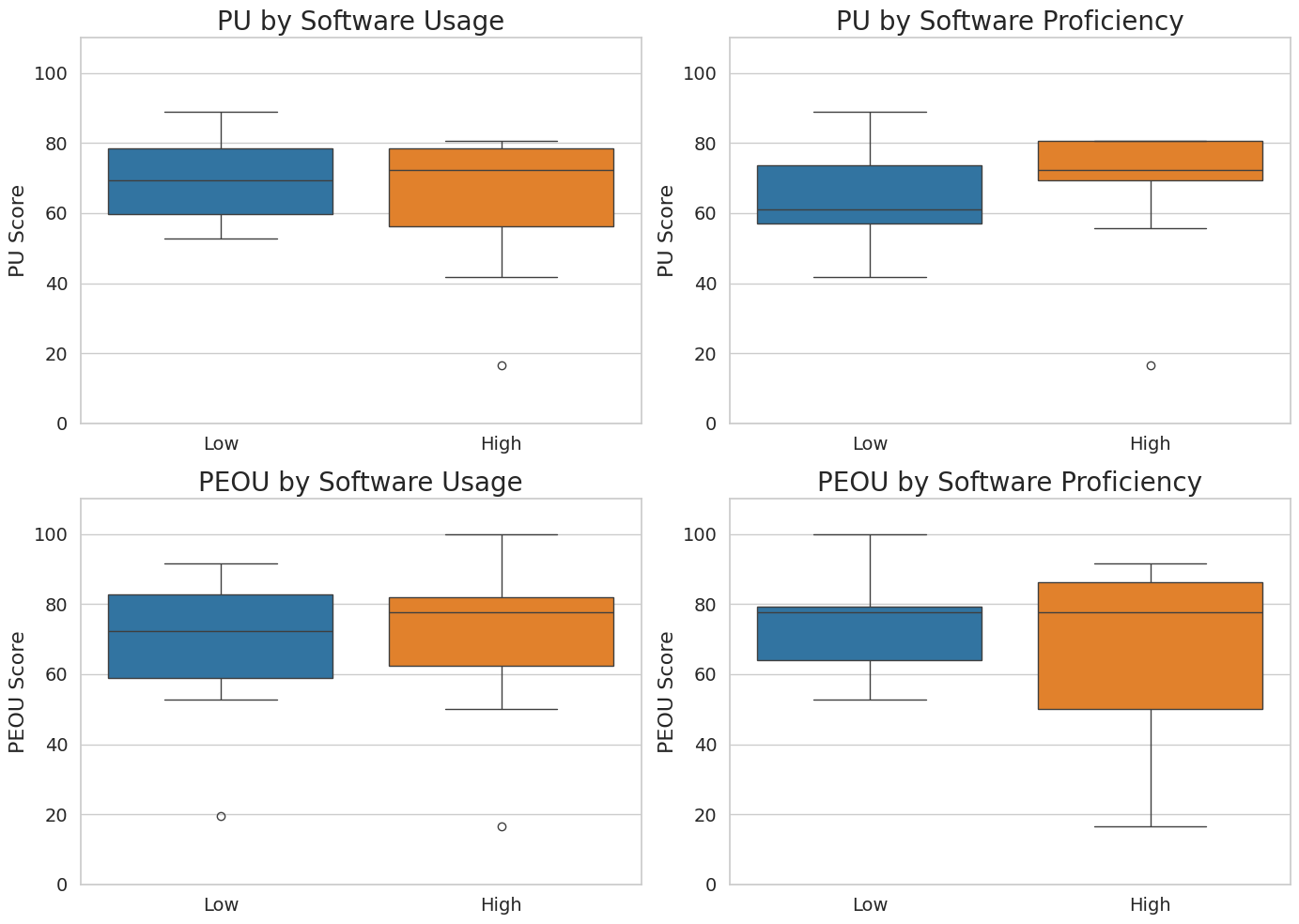}
    \caption{Perceived Usefulness and Perceived Ease of Use by Low (Under 1 hour, 1-3 hours) vs. High (3-5 hours, 6-8 hours) software usage and Low (Basic, Intermediate) vs. High (Advanced, Expert) software proficiency (see \cref{fig:software-usage})}
    \label{fig:tam_by_su}
        \Description{Box plots showing perceived Usefulness and Perceived Ease of Use by Low (Under 1 hour, 1-3 hours) vs. High (3-5 hours, 6-8 hours) software usage and Low (Basic, Intermediate) vs. High (Advanced, Expert) software proficiency (see figure 3)}
\end{figure}

\section{Study Task Text}
\label{app:task}

Please read these instructions thoroughly before starting the task:

Imagine you are a researcher investigating geological structures. 

Connect the Digital Lab Book (DLB) to the Virtual Fieldwork tool (VFT). 
Get an initial overview of the available virtual fieldwork locations and models. 
Think about what you would like to investigate and record these initial thoughts using the DLB, e.g. in the Notes component.

Decide on a research question or research goal and note this down. 
Then use the VFT to investigate your research goal e.g. through measurements or placing markers at interesting locations. 
Use the DLB to thoroughly record your process and organize your findings (provenance graph, mind map, annotations, notes).

Imagine that 
\begin{itemize}
    \item[a)] a colleague will receive (only) your DLB recording to continue and revisit your investigation without further information.
    \item[b)] it will be published as supplementary material alongside a research paper and should enable reviewers to validate your work and understand your reasoning and investigation process. 
\end{itemize}

There is no correct solution and no given goal to this task except what you decide to investigate.

You have roughly 30 minutes after which you should use the DLB to write down your conclusion and findings from this session and export the provenance repository.

\section{Questionnaire}
\label{app:tam}
The modifications were done because participants used the DLB in an artificial scenario not representative of their research work. 
In items referring to "job" we included "job and research".
For the first six items (measuring perceived usefulness) we used the wording from the
original TAM questionnaire e.g. asking participantswhether using the DLB "would increase" their productivity in their job and research.
Lewis's version (2019) instead asks participants whether the given product "increases" their productivity.
For the perceived ease of use, however, we use the past tense wording, asking e.g. whether participants found it easy to get the DLB to do what they wanted it to do, referring to the study session.
\renewcommand{\thefigure}{C.\arabic{figure}}
\setcounter{figure}{0}  
\begin{figure*}[!htbp]
    \centering
    \includegraphics[width=1\linewidth]{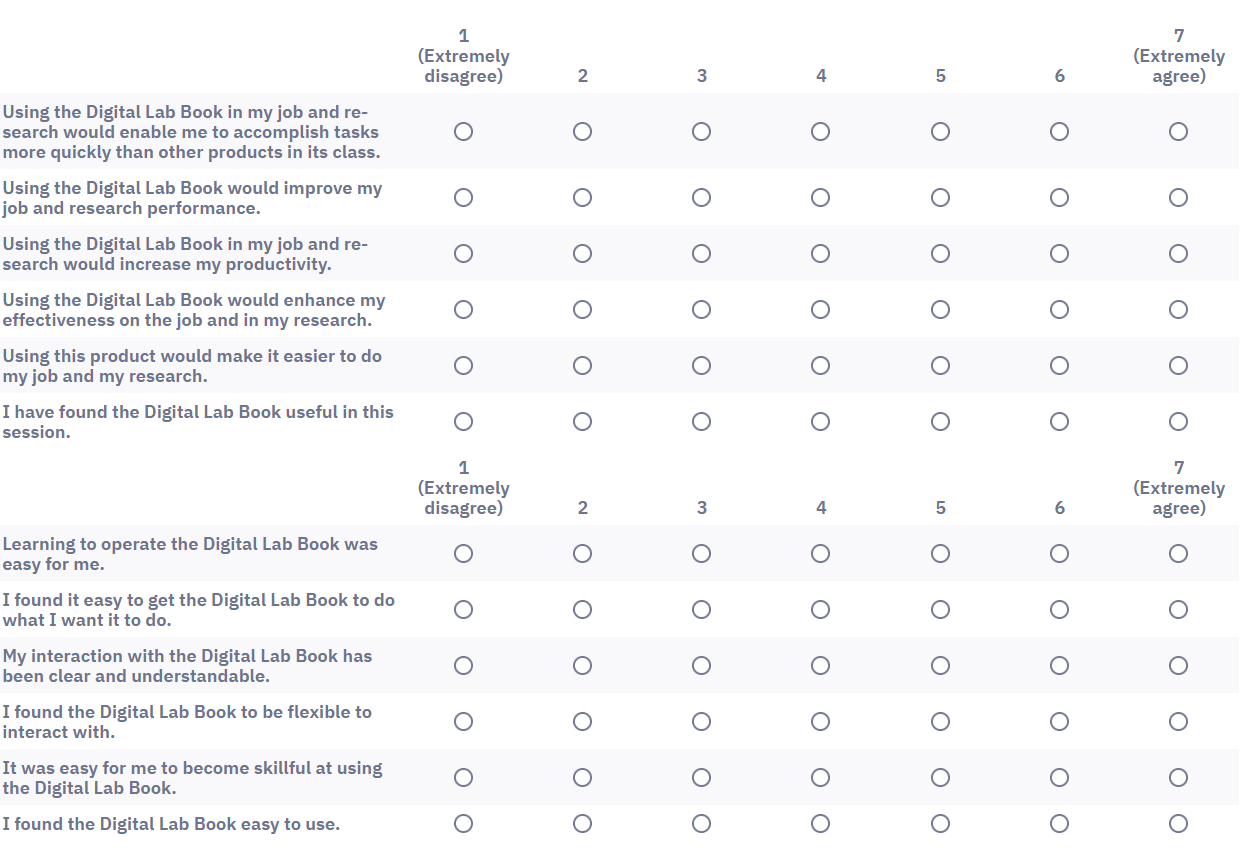}
    \caption{Technology Acceptance Model Questionnaire. Items 1-6 result in the perceived usefulness and items 7-12 the perceived ease of use}
    \label{fig:tam}
    \Description{Intermediate}
\end{figure*}

\end{document}